\documentstyle[12pt,epsf]{article} 
\newcommand{\be}{\begin{equation}}
\newcommand{\en}{\end{equation}}

\begin{document}

\vglue 1truecm
  
\vfil
\centerline{\large\bf Center vortices of Yang-Mills theory } 
\centerline{\large\bf at finite temperatures$^*$ } 
  
\bigskip
\centerline{ K.~Langfeld, O.~Tennert, M.~Engelhardt and H.~Reinhardt }
\vspace{1 true cm} 
\centerline{ Institut f\"ur Theoretische Physik, Universit\"at 
   T\"ubingen }
\centerline{D--72076 T\"ubingen, Germany}
  
\vfil
\begin{abstract}

Recent lattice calculations performed at zero temperature and in the 
maximal center gauge indicate that quark confinement can be understood 
in this gauge as due to fluctuations in the number of magnetic 
vortices piercing a given Wilson loop. This development has
led to a revival of the vortex condensation theory of confinement. 
For a SU(2) gauge group, we show that also at finite temperatures, 
center vortices are the relevant collective infrared degrees of freedom
determining the long-range static quark potential; in particular,
their dynamics reflect the transition to the deconfining phase.

\end{abstract}

\vfil
\hrule width 5truecm
\vskip .2truecm
\begin{quote} 
$^*$ Supported in part by DFG under contract Re 856/1--3. 
\end{quote}
\eject

{\bf Introduction.}
One of the most intriguing prospects of strong interaction physics
is the expected existence of a deconfined phase above a certain
transition temperature. While strongly interacting matter, i.e.
matter carrying a color quantum number, has to date only been
observed in the form of color-singlet bound states called hadrons,
the deconfined phase is characterized by the possibility of
colored constituents propagating freely over distances large
compared with hadronic sizes. Upcoming experiments at RHIC
and LHC are expected to probe the deconfined regime, and thus
efforts to establish an adequately detailed theoretical picture
of the deconfinement phase transition and the deconfined phase
acquire a measure of urgency.

\vskip 0.3cm
The investigation of the deconfinement transition is irrevocably
tied to the question of the mechanism leading to confinement in
the low-temperature phase of QCD, the gauge theory of the
strong interactions. It should be noted that
in different gauges, different characteristic
infrared collective degrees of freedom dominate the physics of
confinement, and the convenience of a gauge hinges on the ease
with which these infrared degrees of freedom can be identified
and isolated. Examples of gauges which are in this sense convenient
are the maximal Abelian gauge, which displays monopole dominance
and thus underpins the dual superconductor picture of confinement
\cite{tho76}-\cite{poli97}, and the maximal center gauges 
\cite{deb96}-\cite{corr}, which display center dominance and lead to 
the center vortex picture of confinement \cite{deb96}-\cite{tom93}. 
It is this latter picture we concentrate on in this letter.

\vskip 0.3cm
Maximal center gauges are defined in the lattice formulation of
Yang-Mills theory by the requirement to choose link variables on the
lattice as close to center elements of the gauge group as the gauge
freedom will allow. There are many slightly differing ways of
implementing this general idea; in this work, the specific prescription
called ``direct maximal center gauge'' was adopted, for details
see e.g.~\cite{deb97}. Furthermore, in what follows, we will be concerned 
with an SU(2) gauge group, the center of which consists of the 
elements $\pm 1$. The maximal center gauges have the following
convenient property: If one replaces the gauge-fixed link variables
with the center elements closest to them (so-called center projection),
then a Monte Carlo calculation of the long-range zero-temperature
static quark potential via the Wilson loop yields virtually the same
result as when using the full unprojected link variables \cite{deb96}. 
This empirical observation is called center dominance. Note that in both
cases, the weight in the Monte Carlo calculation is the full Yang-Mills
action; only the observable, i.e. the Wilson loop, is sampled using
either the full or the projected configurations. Center dominance
is interpreted as meaning that the maximal center gauges successfully
concentrate the information relevant for confinement on the center
part of the link variables, thereby rendering the projection procedure
of discarding the residual deviations from the center elements
inconsequential.

\vskip 0.3cm
In order to further physically interpret the center projected
link configurations, one can connect them with a notion of vortices as
follows. After center projection, one is left with lattice links 
associated with center elements $\pm 1$ of the gauge group SU(2).
If the links bordering any given plaquette on the lattice multiply
to $-1$, then a vortex is said to pierce that plaquette. The vortices
defined in this way form closed two-dimensional surfaces in a 
four-dimensional space-time lattice, or closed (magnetic flux) lines
on the three-dimensional lattice describing, say, one time slice.
An important property of these vortices is that they contribute a
factor $-1$ to any Wilson loop whenever they pierce its minimal area;
in this way, they produce the same value for the Wilson loop as one
obtains by multiplying the original center-projected links making
up the loop (this is simply Stokes' theorem). If a Wilson loop area is
pierced by vortices an even number of times, then it takes the 
value $+1$; if a Wilson loop area is pierced by vortices an odd number of 
times, then it takes the value $-1$. Thus, if the distribution 
of vortices in space-time fluctuates sufficiently randomly, then a 
strong cancellation will occur between configurations where Wilson loops
are linked by an even or by an odd number of vortices; an ever better
cancellation as the area of the Wilson loop increases leads to an area
law decay of the Wilson loop expectation value, which implies
a linear static quark potential. This is the essence of the random 
vortex picture of confinement. After having been proposed as early
as 1978 \cite{thov}-\cite{cornwall}, glimpses of such vortex 
configurations were afforded by the Copenhagen ``spaghetti'' vacuum 
\cite{ole}, and several investigations were devoted to identifying 
vortices on the lattice, leading to different definitions of vortices
\cite{mack},\cite{tom93},\cite{deb96}. The recent efforts in the 
framework of the maximal center gauges outlined above 
\cite{deb96}-\cite{corr} have given new impetus to this 
particular description of confinement, foremost due to the
verification of center dominance \cite{deb96}. Subsequently, further 
properties of center vortices have been investigated, e.g. the 
connection to the monopoles of the maximal Abelian gauge \cite{deb96b}, 
and the observation that the density and binary correlations of these 
vortices display the correct renormalization group scaling properties 
\cite{la97},\cite{corr}. This means that they survive the continuum 
limit and represent physical objects rather than lattice artifacts.

\vskip 0.3cm
Having identified such a viable picture of confinement, it is natural 
to confront it with the question of the deconfinement transition.
In this letter, we verify that the center vortex picture makes sense
also at finite temperature, since center dominance persists. 
Also the transition to the deconfined phase with a vanishing
string tension is correctly reproduced. We further explore 
properties of vortices as the temperature is raised across the 
deconfinement transition and we cast the deconfinement order 
parameter into the vortex language.

\begin{figure}[t]
\centerline{ 
\epsfxsize=9cm
\epsffile{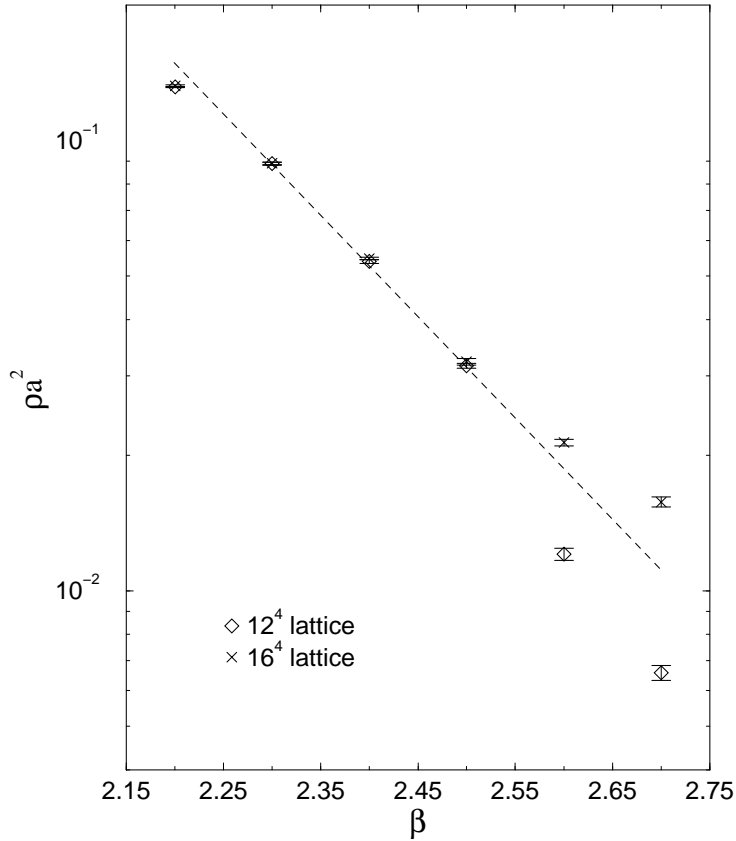}
}
\caption{Scaling of the planar density of vortex intersection points
with a given space-time plane.}
\label{fig:e1} 
\end{figure}
\vskip 0.3cm
{\bf Zero temperature vortex properties.} 
For later reference and in order to enable the reader to estimate
the accuracy which is achieved in our numerical calculations, 
we briefly summarize our zero temperature results. Our numerical data
were obtained using lattices of size $12^4$ and $16^4$. 
The string tension is extracted from center-projected links by 
calculating $n\times n$ Creutz ratios, $n=3,4,5$. One observes that 
the results for the Creutz ratios for different $n$ agree already for
these small $n$, indicating that the Coulomb part of the heavy quark 
potential is absent when the latter is evaluated using center-projected 
links \cite{deb96}. The Creutz ratios allow us to extract
$\sigma a^2(\beta=2.3) = 0.12(3) $. The $\beta $-dependence of 
$\sigma a^2$ is compatible with one--loop perturbative scaling. 
Two-loop effects are small compared with the statistical error of 
our numerical data. Furthermore, for the lattices used ($12^4$ and $16^4$),
the finite size dependence of dimensionless quantities was
negligible compared with the statistical errors. However, as is well known,
considerable finite size effects persist in the extraction of
the mass scale, as e.g. encoded in the lattice spacing, as a function
of $\beta $. In the large scale analysis \cite{fing93}, an interpolation
of the numerical data with the help of one--loop scaling to the
renormalization point $\beta = 2.3$ reveals $\sigma a^2 = 0.136 $ for a
$10^4$ lattice, $\sigma a^2 = 0.121 $ for a $16^4$ lattice, while
$\sigma a^2 = 0.107 $ is found for a $32^4$ lattice. Compatible with this,
we use throughout this letter $\sigma a^2(\beta=2.3) = 0.12 $ 
and $\sigma = (440 \, \hbox{MeV} ) ^2 $ as reference scale for 
assigning physical units to measured quantities. When finite size
uncertainties are mentioned below, this always refers to the 
aforementioned difficulty in defining the mass scale at a given $\beta $.

\vskip 0.3cm
In order to extract vortex properties, we fix the configurations to
the direct maximal center gauge by using the procedure described 
in~\cite{deb97}. To eliminate Gribov copies, we perform random 
gauge transformations and repeat minimizing the gauge fixing functional. 
It turns out that three of these runs are sufficient at zero 
temperature~\cite{deb97}. For temperatures close to the transition 
point, we observe an increasing influence of the Gribov copies 
making six gauge fixing iterations necessary. The scale dependence 
of the vortex density is shown in figure \ref{fig:e1}. 
We find a perfect scaling of the planar density $\rho $ of vortex
intersection points with a given space-time plane
in the range $\beta \in [2.25,2.55]$ for lattice sizes $12^4, 16^4$. 
We finally extract $ \rho = 3.6 \, \pm 0.2 \, \hbox{fm}^{-2}$. The large 
error bar in $\rho $ results from the uncertainty in the magnitude of finite 
size effects. The above value, which is larger than the one quoted 
in~\cite{la97}, is now in agreement with the one quoted in \cite{deb97}
within the error bars. The faulty estimate for $\rho $ given
in~\cite{la97} is due to the fact that the Coulomb part of the full zero
temperature static quark potential was underestimated in~\cite{la97}, 
which subsequently led to an overestimate of the reference scale.
As a consequence, mass scales in~\cite{la97}, and also in~\cite{corr},
should be rescaled upwards by a factor $1.35$. Note also that the new 
corrected value of $\sigma / \rho = 1.4 \pm 0.1 $ as opposed to the old 
value $\sigma / \rho = 2.5 $ implies that a model of random intersection
points of vortices with a two-dimensional space-time plane \cite{corr}
overestimates, rather than underestimates, the string tension as a
function of the planar density of points $\rho $. Such a model \cite{corr}
leads to a value of $\sigma / \rho = 2$. However, this does not affect 
the motivation for the lattice measurements in \cite{corr} nor their
validity, up to the aforementioned trivial rescaling of mass scales.
In \cite{corr}, correlations of an attractive type were observed
between intersection points of vortices with a two-dimensional 
space-time plane. It seems plausible that such correlations curtail the
randomness of the distribution of the points and thus reduce the string
tension. Indeed, further below it will become clear that it is a pairing of
the vortex intersection points which ultimately leads to deconfinement.
However, one should bear in mind that the connection between correlators
such as measured in \cite{corr} and the string tension is very indirect
and one can not in general predict the behavior of the string tension
from these correlators alone.

\vskip 0.3cm
{\bf Center dominance at finite temperatures.}
Finite temperature Yang-Mills theory can be cast into a
path integral formulation on a space-time manifold of finite
(Euclidean) time extension $1/T$, where $T$ is the temperature.
Gauge fields on the lattice obey periodic boundary conditions in 
the time direction. The static quark potential $V(r)$ is calculated 
from the correlator of two Polyakov loops \cite{sve},
\begin{eqnarray}
\left\langle L(\vec{x}) \, L(\vec{y}) \right\rangle
& \propto & \exp \left\{ - V(r) /T \right\} , 
\label{vdeff} \\
L(\vec{x}) & = & {\cal P} \exp \left( i \int _{0}^{1/T} A_0(t, \vec{x})
\, dt \right) ,
\end{eqnarray}
where $A_\mu (x) $ is the gauge potential, ${\cal P}$ denotes path 
ordering, and $r = \vert \vec{x} - \vec{y}\vert$. 

\vskip 0.3cm
Note that, since center elements of the gauge group commute, the
Polyakov loop correlator evaluated with center-projected
configurations is equal to the Wilson loop of identical spatial
width and extending along the entire time direction. Therefore,
center vortices contribute to the static quark potential at
finite temperatures in the same way as they do at zero temperature
(i.e. Stokes' theorem applies in the same way). The question to be
settled is whether, at finite temperature, vortices still provide
the entire long-range static quark potential (center dominance)
or whether other effects become important. This would mean that
center vortices lose their role as relevant infrared collective
degrees of freedom. 

\vskip 0.3cm
We have tested this empirically employing Monte Carlo measurements.
On the one hand, we have evaluated Polyakov loop correlators using
center-projected links on a $12^3 \times N_t $ lattice with 
$N_t =5,6,7$ for different $\beta \in [2.26 \ldots 2.4 ]$, 
i.e. different inverse temperatures $1/T = N_t a(\beta )$ 
were achieved by varying $\beta $. The $\beta $-dependence 
of the lattice spacing $a$ was extracted from zero 
temperature string tension measurements (see previous section).
While these measurements are thus fraught with sizeable
uncertainties due to finite size effects in $a(\beta ) $, the 
statistical fluctuations still turn out to be the dominant source of 
error.

\vskip 0.3cm
The center-projected string tension $\sigma_{t} $ as a function
of temperature was extracted from the Polyakov loop correlators
by fitting a linear law to the potential $V(r)$ of eq. (\ref{vdeff}).
A Coulomb term is not necessary, since, as will be illustrated by 
an example below, center projection removes the perturbative Coulomb part 
from $V(r)$ just as at zero temperature \cite{deb96}. 
In addition, however, it should be kept in mind that the static
quark potential at finite temperatures in general also contains
a logarithmic dependence on the separation \cite{kacz}. Thus,
fitting a purely linear law to the potential strictly speaking does
not yield the coefficient of the linear term, which by definition
constitutes the string tension; instead, one obtains an effective 
``string tension'' which provides a good parametrization of the
long-range static quark potential in the limited range of separations
accessible to lattice experiments. Since the accuracy of our
measurements is limited, we cannot separate the linear and 
logarithmic parts of the potential, and we thus quote instead
the effective string tension in the sense explained above.
This quantity in full Yang-Mills theory is known to behave
as follows. It retains its zero-temperature value to within 
approximately 10\% up to the temperature $0.8 \, T_c $, where
$T_c $ is the deconfinement phase transition temperature, and then
quickly drops to zero. For the SU(2) theory, we have only been
able to find some rather sparse older data to substantiate this
\cite{altst}; on the other hand, for the SU(3) case, which should
behave qualitatively in the same manner, new high precision measurements
exist \cite{carle} which corroborate the aforementioned behavior of 
the effective string tension.

\begin{figure}[t]
\centerline{ 
\epsfxsize=7cm
\epsffile{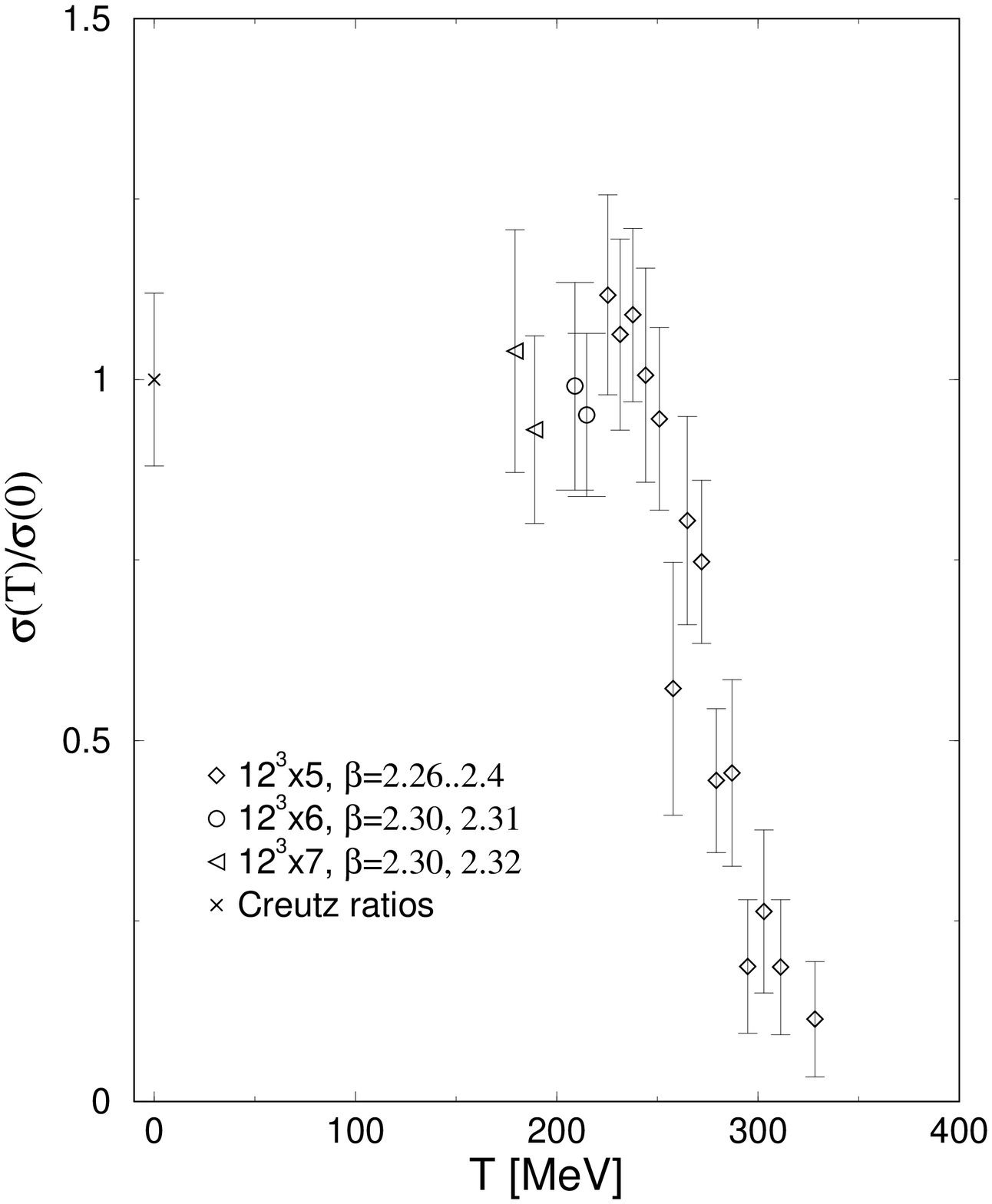}
\hspace{1cm}
\epsfxsize=7cm
\epsffile{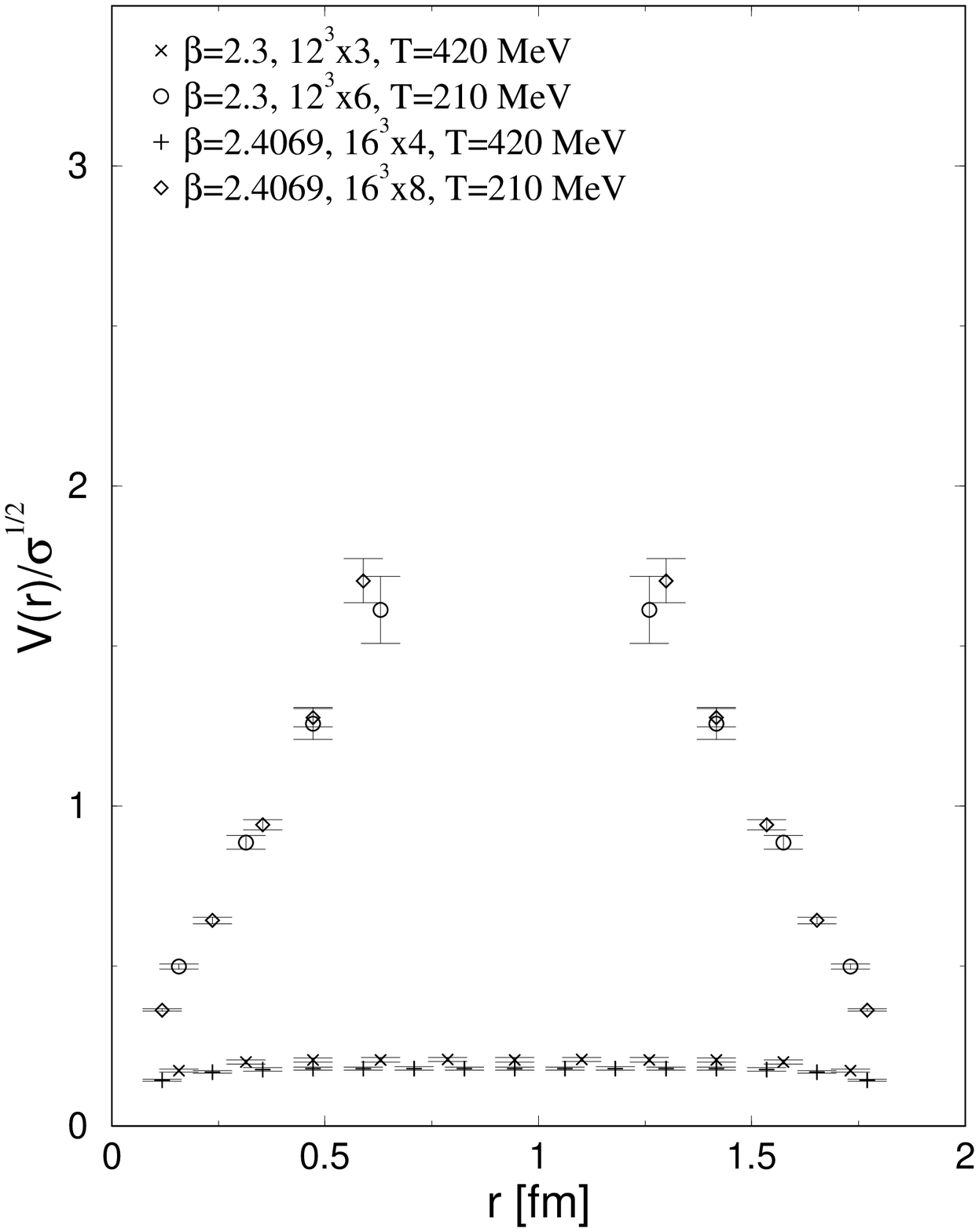}
}
\caption{Left: The string tension $\sigma _t$ calculated from
center-projected links as function of the temperature $T$;
the four lowest-temperature points on the $12^3 \times 5$ lattice
are measured at $\beta < 2.3$ and thus may already be subject
to systematic scaling violations. The zero temperature measurement
is included to give an indication of the error bars.
Right: The static quark potential at two sample temperatures, as extracted 
from Polyakov loop correlators with center-projected links, in units
of the zero-temperature string tension.}
\label{fig:1} 
\end{figure}

\vskip 0.3cm
Our results for the center-projected case are displayed
in the left-hand plot of Figure \ref{fig:1}. We find that the
center-projected string tension $\sigma _t$ within the error bars
reproduces the behavior of the full string tension quoted
above; thus, we observe center dominance of the long-range part of
the static quark potential at finite temperatures $T< T_c$ within 
the accuracy of our measurement and the limited range of separations 
available.

\vskip 0.3cm
Moreover, the center-projected string tension $\sigma _t$ signals the 
deconfinement phase transition to occur at $T_c = 295 \pm 10$ MeV, 
where the input scale was a zero temperature string tension 
of $\sigma = (440$ MeV$)^2 $. This value of $T_c$ is in good agreement 
with the high precision measurements of~\cite{bal93}. Thus, also the
transition to the deconfined phase with a vanishing string tension
is accurately reproduced by the center-projected theory.

\vskip 0.3cm
For further illustration, we also calculated Polyakov loop 
correlators using in particular $\beta =2.3$ for a $12^3 \times N_t$ 
lattice and $\beta = 2.4069$ for a $16^3 \times N_t$ lattice while
varying the number of lattice points $N_t $ in time direction in order
to obtain different inverse temperatures $1/T = N_t a(\beta )$.
The $\beta $ values and lattice sizes in both cases correspond 
to a physical size of the spatial cube of 
$L_s = N_s a(\beta ) \approx 1.9 \, $fm, while the lattice spacing
changes by a factor $a(\beta=2.3)/a(\beta=2.4069) = 4/3 $. 
This particular choice of lattice sizes and $\beta $--values therefore 
allows to test the scaling behavior of the lattice observables 
with vastly diminished fluctuations due to finite size effects.
Of course, the overall uncertainty in the mass scale remains.
In the case of Polyakov loop correlators, this does not
turn out to be crucial; the statistical fluctuations are the main 
source of error. However, when measuring vortex densities (see next
section), this method substantially improves the observed scaling
properties, since the statistical errors are not as dominant.

\vskip 0.3cm
Using the Monte Carlo results for center-projected Polyakov loop
correlators on these lattices, the static quark potential as a 
function of the distance at diverse temperatures both below and 
above the deconfinement transition was extracted. Two examples,
corresponding to $N_t = 3,6$ for $\beta =2.3$ and
$N_t = 4,8$ for $\beta =2.4069$,
are given in the right-hand plot in Figure \ref{fig:1}. 
For $T< T_c$, the potential rises linearly even at small distances;
as observed previously for the case of zero temperature \cite{deb96},
center projection removes the short range Coulomb interaction.

\begin{figure}[t]
\centerline{
\epsfxsize=7cm
\epsffile{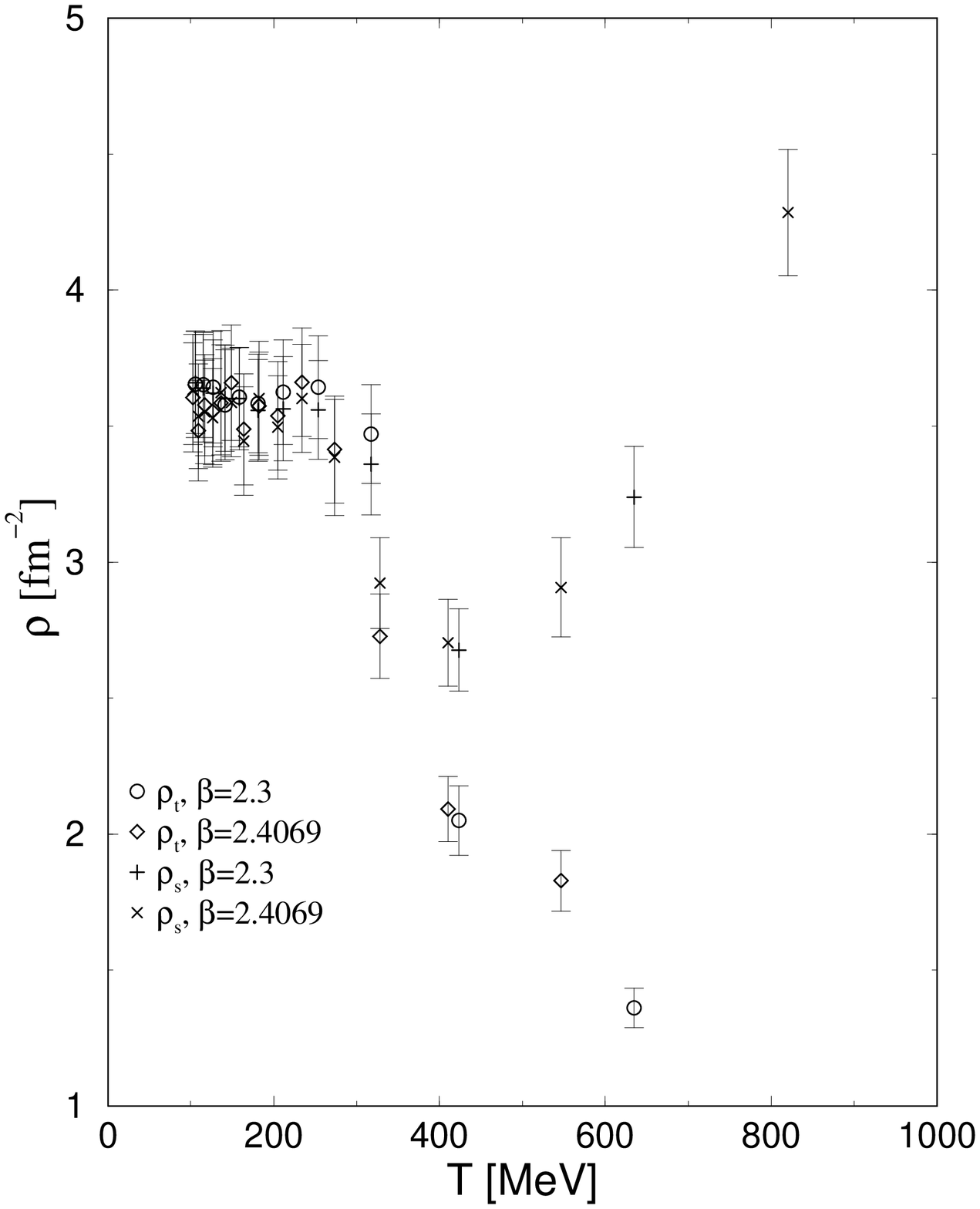}
\hspace{1cm} 
\epsfxsize=7cm
\epsffile{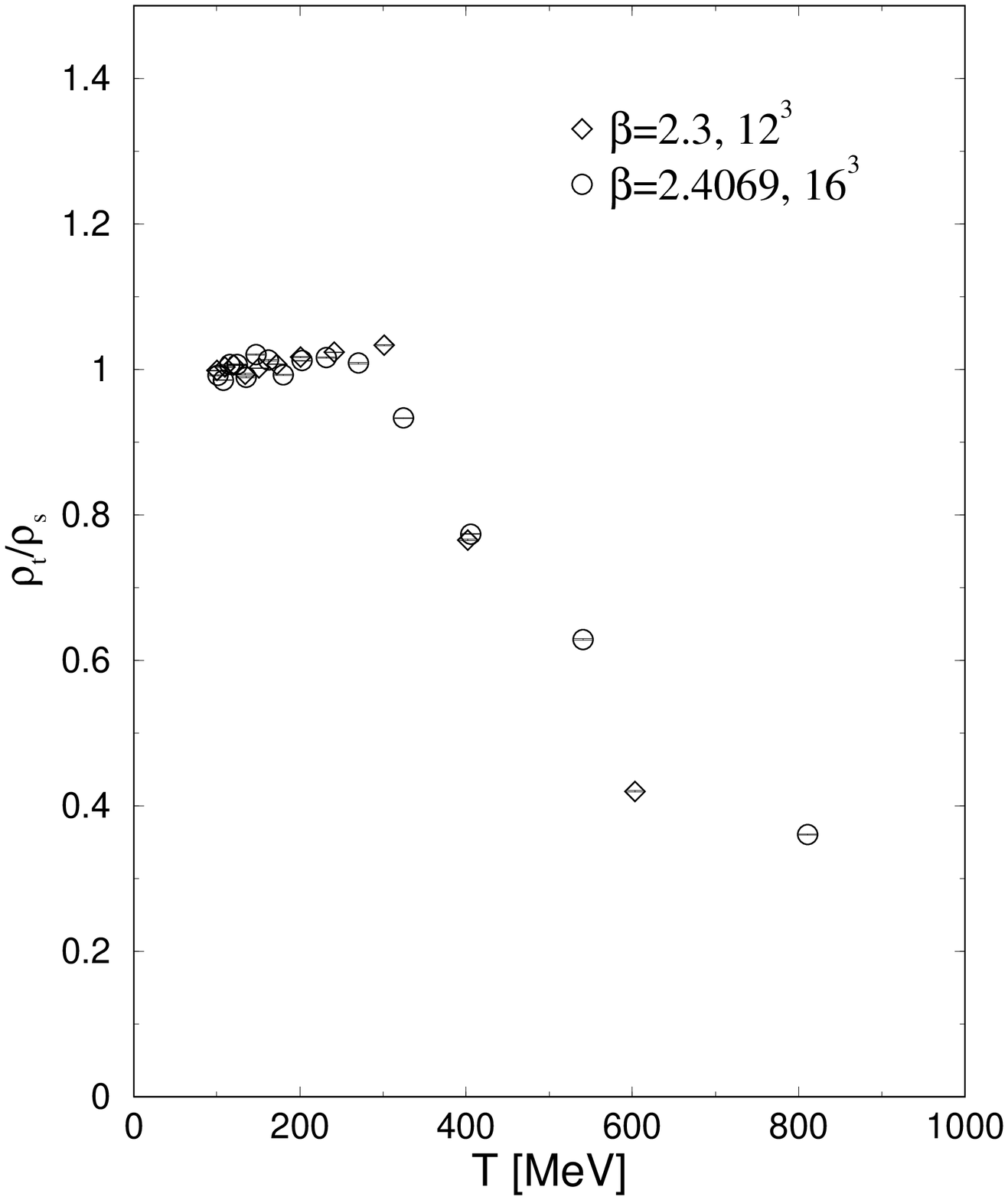}
}
\caption{The planar densities $\rho _t$ and $\rho _s$ of points at
which vortices pierce time-like and space-like planes, respectively.}
\label{fig:1b}
\end{figure}

\vskip 0.3cm
{\bf Vortex polarisation.} 
Given this success of the vortex picture,
the question of the nature of the deconfinement transition in this
picture poses itself. One of the simplest hypotheses would appear to
be the following: While the center vortices are by construction one
lattice spacing thick, they represent smooth configurations in the
original gauge fields before the gauge fixing and center projection
procedure is applied, with a physical thickness in the continuum 
limit \cite{deb96},\cite{corr}. Possibly vortices running 
perpendicularly to the time direction are simply too thick to fit into 
the space-time manifold of time extension $1/T$ for $T>T_c $. This would 
mean that the planar density of points at which vortices pierce the area 
spanned by the two Polyakov loops entering the loop correlator
vanishes. Vanishing density of such points precludes fluctuations in
the number of such points, making an area law decay of the Polyakov
loop correlator impossible. In order to test this scenario, we have
measured the planar density $\rho_{t} $ of points at which vortices
pierce an area extending in the time and one space direction, for
different temperatures. In order to reduce fluctuations due to
finite size effects (see previous section),
these measurements were done using $\beta =2.3$ for a $12^3 \times N_t$
lattice and $\beta = 2.4069$ for a $16^3 \times N_t$ lattice while
varying the number of lattice points $N_t $ in time direction in order
to obtain different inverse temperatures $1/T = N_t a(\beta )$.
The result for the vortex density is shown in Figure \ref{fig:1b}.

\vskip 0.3cm
Evidently, while $\rho_{t} $ experiences a drop as one increases the
temperature past $T_c $, the behavior of $\rho_{t} $ is smooth and
at $T\approx 2 T_c $, $\rho_{t} $ still retains roughly a third of its
zero temperature value. Thus, the hypothesis advanced above of
deconfinement being due to a vanishing of $\rho_{t} $ was too
simplistic. There still exists a density of vortices piercing the
area spanned by two Polyakov loops above $T_c $, but the random
character of the distribution of these vortices must disappear in order
to realize deconfinement. We will come back to this presently; before
doing so, note that we have for comparison also measured the density
$\rho_{s} $ of vortices piercing an area extending in two spatial
directions. The ratio $\rho _t / \rho_s$ is also shown in figure 
\ref{fig:1b}. For low temperatures,
$\rho_{t} $ and $\rho_{s} $ coincide, as they must do due to
Euclidean O(4) invariance. At temperatures slightly above $T_c $,
$\rho_{s} $ decreases along with $\rho_{t} $. At higher temperatures,
however, $\rho_{s} $ begins to increase. This seems consistent with
a simple picture of the intersection points making up the density
$\rho_{s} $ still being distributed randomly, which leads to a linear
relation \cite{corr} between the planar density $\rho_{s} $ and the 
corresponding spatial ``string tension'' $\sigma_{s} $ extracted from 
spatial Wilson loops. This spatial string tension in turn is known to 
increase as $\sqrt{ \sigma_{s} } \sim g^2(T) \, T$ for $T \ge 2 T_c $ 
according to dimensional reduction arguments \cite{app81} and their 
verification in lattice experiments \cite{bal93}.
As mentioned above, in the case of the intersection points 
making up the density $\rho_{t} $, this simple random picture must by 
contrast become invalid above $T_c $.

\begin{figure}[t]
\centerline{ 
\epsfxsize=7cm
\epsffile{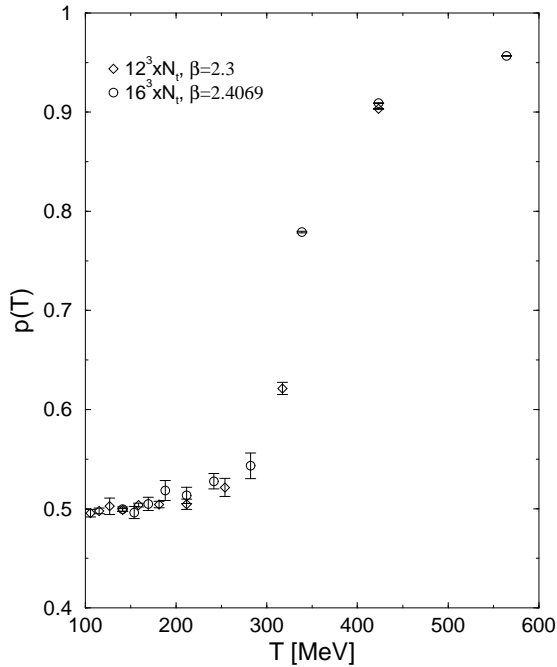}
}
\caption{Fraction of cases in which an area spanned by two Polyakov
loops was pierced an even number of times by vortices.}
\label{fig:2} 
\end{figure}

\vskip 0.3cm
{\bf Pairing of vortex intersection points.}
Coming back to the issue of the vortex description of the deconfinement
transition, it is necessary to inspect more closely the properties of 
vortices piercing the area spanned by the two Polyakov loops entering
a loop correlator. A necessary condition for an area law suppression of 
the loop correlator expectation value is an ever better mutual 
cancellation, as the area spanned by the Polyakov loops is increased, 
between configurations where the area is pierced an even or an odd number 
of times by vortices, respectively. We have therefore measured the
probabilities of these two cases as a function of temperature for an 
area of spatial width $0.9\, $fm. For $\beta =2.3$, this corresponds 
to 6 lattice spacings in the $12^3 \times N_t $ lattice, whereas for 
$\beta =2.4069$, it corresponds to 8 lattice spacings in the 
$16^3 \times N_t $ lattice, i.e. in both 
cases the distance between the Polyakov loops is half the linear extension
of the universe. The result of these Monte Carlo experiments is shown in 
Figure \ref{fig:2}, which displays the fraction $p(T)$ of cases where 
an area specified as above was pierced an even number of times by vortices. 
This quantity exhibits a sharp transition at the deconfinement 
temperature $T_c \approx 295\, $MeV. Note that $p\not\to 1/2$ for large 
areas precludes an area law decay of the Polyakov loop correlator, 
implying deconfinement.

\vskip 0.3cm
The clear signal of the deconfinement transition exhibited by the
above probability $p$ is not surprising, since it constitutes nothing
but a slightly more physical variant of the usual Polyakov loop
order parameter. The center-projected Polyakov loop correlator
is given as follows in terms of the probability $p$ that an even
number of vortices pierces the area spanned by the two Polyakov loops
in question:
\begin{equation}
\left\langle L(\vec{x}) \, L(\vec{y}) \right\rangle =
p\cdot 1 + (1-p)\cdot (-1) = 2p-1 \ .
\label{evodd}
\end{equation}
Thus, $p$ corresponds, up to a rescaling and a shift, to the
Polyakov loop correlator itself. The only difference to the usual
Polyakov loop order parameter, which is related to the free energy
associated with placing a static color source in the thermal medium,
is that the static color source has been given a partner (separated
from it by half the extension of the universe).

\vskip 0.3cm
Therefore, in defining the probability $p$, we have not really introduced
a new, distinct, order parameter for the deconfinement transition. Rather,
the vortex picture simply provides an alternative language for describing
the Polyakov loop correlator, by characterizing the manner in which vortex
intersection points occur on the area spanned by the two Polyakov loops.
The deconfinement transition occurs when the intersection points of
vortices with the area spanned by the Polyakov loops begin to occur
predominantly in pairs of finite separation (above $T_c $), whereas
they occur randomly below $T_c $.

\vskip 0.3cm
{\bf Summary and Outlook.} 
In the present work, we have explored the center vortex content of
Yang-Mills theory at finite temperatures using numerical lattice
methods in the direct maximal center gauge and center projection.
We have observed vortex dominance for the long-renge part of
the static quark potential. 
This establishes the relevance
of center vortices at finite temperatures and thus provides the
necessary basis for further investigation of these degrees of freedom.
\vskip 0.3cm

The planar densities of intersection points of vortices with a given
plane in space-time exhibit an anisotropic behavior as temperature
is raised from the $O(4)$-invariant zero temperature limit.
Around the deconfinement temperature $T_c $, the density $\rho_{t} $
of intersection points with a plane running in the time and one space
direction begins to decrease and retains roughly a third of its zero
temperature magnitude at $T=2T_c $. The density $\rho_{t} $ is relevant
for the behavior of the Polyakov loop correlator, since the area between
two Polyakov loops extends in the time and one space direction. However,
since $\rho_{t} $ does not vanish for $T>T_c $, deconfinement cannot
be simply ascribed to the behavior of $\rho_{t} $, i.e. the partial
polarization of the vortices in time direction. Instead, more detailed
correlations between the intersection points building up the density
$\rho_{t} $ must be present.
\vskip 0.3cm

By contrast, the behavior of the density $\rho_{s} $ of intersection
points with planes running in two spatial directions largely parallels
the known behavior of the spatial string tension $\sigma_{s} $
\cite{app81},\cite{bal93}. After a slight dip around $T_c $, which 
may be a finite-size artefact, $\rho_{s} $ begins to rise with temperature
along with the spatial string tension $\sigma_{s} $.
This is consistent with a simple picture of a random distribution of
the intersection points making up the density $\rho_{s} $. In such a
picture, the associated (spatial) string tension is proportional to
the density \cite{corr}. 
\vskip 0.3cm

Returning to the question of the deconfinement transition, we measured
in more detail the probability that an even or an odd number of
intersection points occurs on the area spanned by two Polyakov loops
at a large distance from one another (half the lattice universe).
This quantity exhibits the deconfinement transition very clearly,
which is not surprising, since (up to a trivial shift) it corresponds to
the value of the Polyakov loop correlator itself, cf. eq. (\ref{evodd}).
This is a slightly more physical realization of the usual Polyakov loop
order parameter, namely, instead of putting one static color source on
the lattice and measuring the associated free energy, we give the
color source a static partner at the edge of the universe.
Thus, we have not really defined a new order parameter, but the
vortex picture provides an alternative language for describing the
Polyakov loop correlator: The deconfinement transition occurs when
the intersection points of vortices with the area spanned by the
Polyakov loops begin to occur predominantly in pairs of finite separation
(above $T_c $), whereas they occur randomly below $T_c $. We conjecture 
this behavior to be connected with a percolation transition in the 
vortex surfaces: Since the vortex surfaces are closed, intersection 
points of these surfaces with the entire plane containing two Polyakov 
loops must always occur in pairs. What determines the behavior of the
Polyakov loop correlator is whether the members of a pair are
only separated by a distance small compared with the separation of
the Polyakov loops, such that most pairs lie either
entirely inside or outside the area spanned by the Polyakov loops, or
whether the members of a pair can occur arbitrarily far apart, such
that the distribution of points is essentially random. In terms of
the vortex surfaces, this translates into the question whether typical
vortex clusters have a finite extension, or whether they form networks
extending over the entire universe, i.e. percolate. This leads to
the conjecture that the confining phase is a phase in which the
vortices percolate throughout space-time, while in the deconfining phase, 
they cease to do so. We plan to report on a detailed examination of the
vortex percolation properties in an upcoming publication.

\vspace{1cm}
{\bf Acknowledgements.} M.E. acknowledges a useful discussion with
F.Karsch on the static quark potential at finite temperatures.
This work was supported in part by DFG under contract Re 856/1--3.


\begin{thebibliography}{sch90}
\bibitem{tho76} S.~Mandelstam, Phys. Rep. {\bf 23C} (1976) 245; \\
G.~'t~Hooft, Nucl. Phys. {\bf B190} (1981) 455.
\bibitem{wiese} A.~Kronfeld, M.~Laursen, G.~Schierholz and U.-J.~Wiese,
Nucl. Phys. {\bf B293} (1987) 461.
\bibitem{suz90} T.~Suzuki and I.~Yotsuyanagi, Phys. Rev. {\bf D 42} 
(1990) 4257.
\bibitem{poli97} M.~I.~Polikarpov, 
Nucl. Phys. Proc. Suppl. {\bf 53} (1997) 134.
\bibitem{deb96} L.~Del Debbio, M.~Faber, J.~Greensite and
\v{S}.~Olejn{\'\i}k, Phys. Rev. {\bf D 55} (1997) 2298.
\bibitem{deb96b} L.~Del Debbio, M.~Faber, J.~Greensite and
\v{S}.~Olejn{\'\i}k, talk presented at the NATO Advanced Research
Workshop on Theoretical Physics: New Developments in Quantum
Field Theory, Zakopane, Poland, 14-20 June 1997, hep-lat/9708023.
\bibitem{deb97} L.~Del Debbio, M.~Faber, J.~Giedt,
J.~Greensite and \v{S}.~Olejn{\'\i}k, 
Phys. Rev. {\bf D 58} (1998) 094501, hep-lat/9801027. 
\bibitem{la97} K.~Langfeld, H.~Reinhardt and O.~Tennert,
Phys. Lett. {\bf B419} (1998) 317.
\bibitem{corr} M.~Engelhardt, K.~Langfeld, H.~Reinhardt and O.~Tennert,
Phys. Lett. {\bf B431} (1998) 141.
\bibitem{fing93} J.~Fingberg, U.~Heller and F.~Karsch, 
Nucl. Phys. {\bf B392} (1993) 493.
\bibitem{thov} G.~'t~Hooft, Nucl. Phys. {\bf B138} (1978) 1.
\bibitem{aha78} Y.~Aharonov, A.~Casher and S.~Yankielowicz,
Nucl. Phys. {\bf B146} (1978) 256.
\bibitem{cornwall} J.~M.~Cornwall, Nucl. Phys. {\bf B157} (1979) 392; \\
J.~M.~Cornwall, hep-th/9712248.
\bibitem{ole} J.~Ambj{\o}rn and P.~Olesen, Nucl. Phys. {\bf B170}
[FS1] (1980) 265; \\ P.~Olesen, Nucl. Phys. {\bf B200} [FS4] (1982) 381.
\bibitem{mack} G.~Mack, in ``Recent developments in gauge theories'', 
eds. G.~'t~Hooft et al. (Plenum, New York, 1980); \\
G.~Mack and E.~Pietarinen, Nucl. Phys. {\bf B205} [FS5] (1982) 141.
\bibitem{tom93} E.~T.~Tomboulis, Phys. Lett. {\bf B303} (1993) 103; \\ 
T.~G.~Kov\'acs and E.~T.~Tomboulis, hep-lat/9711009.
\bibitem{sve} B.~Svetitsky, Phys. Rep. {\bf 132} (1986) 1.
(1982) 423.
\bibitem{kacz} O.~Kaczmarek, Diploma thesis, Bielefeld 1997.
\bibitem{altst} F.~Karsch and C.~B.~Lang, Phys. Lett. {\bf B138}
(1984) 176.
\bibitem{carle} C.~DeTar, O.~Kaczmarek, F.~Karsch and E.~Laermann, 
Phys. Rev. {\bf D 59} (1999) 031501.
\bibitem{bal93} G.~S.~Bali, J.~Fingberg, U.~M.~Heller, F.~Karsch
and K.~Schilling, Phys. Rev. Lett. {\bf 71} (1993) 3059.
\bibitem{app81} T.~Appelquist and R.~D.~Pisarski, Phys. Rev. {\bf D 23} 
(1981) 2305.
\end{thebibliography}
\end{document}